\documentclass[prl,twocolumn,showpacs,preprintnumbers,aps]{revtex4}

\usepackage{amsmath,amssymb,amscd}
\usepackage{mathrsfs}

\usepackage{bm}

\begin{document}

\preprint{xxx/xxx-xxx}

\title{Energy Conservation in Flat FRW Cosmology}
\author{Steven Maxson}
\affiliation{ University of Colorado Denver \\  Denver, Colorado 80217}
\email{steven.maxson@ucdenver.edu}

\date{\today}                                        

\begin{abstract}
The consequence of energy conservation in the flat Friedmannn-Robertson-Walker (FRW) cosmology is a strictly positive accelerating expansion. A mechanism is proposed for this expansion due to the effect of the attractive (negative) gravitational potential of matter as it is being included within the expanding horizon, and the offsetting work of metric expansion, which takes place at sub-luminal speed. In our semi-classical treatment, we deal with a quintic as the equation for the scale parameter. Implications for modeling the earliest parts of the primordial expansion are discussed.
\end{abstract}


\maketitle

\section{\label{sec:intro} Preliminaries}

At cosmological scales, spacetime seems to be well modeled in the current era by the flat Friedmann--Robertson--Walker (FRW) cosmology. We will assume the flat FRW cosmology as a standard result, and its implications such as the Friedmann equation without elaboration~\cite{note1}. Thus, when the metric is expressed in the form $diag (-1, a^2(t),a^2(t),a^2(t) )$, the Friedmann equation in the flat case 
\begin{equation}
\left( \frac{\dot a}{a} \right) = \frac{8 \pi G}{3} \rho  \,  . 
    \label{eq:Friedmann}
\end{equation}
in the case of conservation of energy yields
\begin{equation}
0 = \dot\rho = \frac{3}{8\pi G} \left\{ 2 \left( \frac{\dot a}{a} \right) \left[ 
     \frac{\ddot a}{a} -\left( \frac{\dot a}{a} \right)^2  \right] \right\} \, .
     \label{eq:Econs1}
\end{equation}           
which implies
\begin{equation}
\frac{\ddot a}{a}  = \left( \frac{\dot{a}}{a} \right)^2
     \label{eq:addot}
\end{equation}
Given the observation of a cosmological red-shift, we conclude that $\ddot a > 0$, the cosmological acceleration must have been strictly positive if energy has been conserved during the flat FRW era.
(We ignore any formally possible inverting cosmology in which $a<0$ as physically uninteresting, since then one merely deduces accelerating expansion in the inverted sense.) The accelerating expansion of this model is consistent with observation~\cite{perlmutter,riess1}.

The conservation law
\begin{equation}
\dot\rho = - \frac{3 \dot a}{a} (\rho + p) \, , 
     \label{eq:conslaw}
\end{equation}
implies that $p = - \rho$ in the case of energy conservation $\dot\rho = 0$. This consequence of energy conservation is consistent with observation~\cite{riess2}.

The above are results relevant to later discussions, not new observations. The interesting issues are how and why can energy be conserved and what is the mechanism driving the observed accelerating expansion. We will address these key issues in the next section, but the key motivation will be:
\begin{itemize}
\item{Idealization of the universe as a thermodynamically closed system.}
\item{The observation that, as the horizon of an observer expands at the speed of light, it is ``discovering'' new matter at a finite distance, and the gravitational attraction will result in a negative gravitational potential energy equivalent to bringing the newly ``discovered'' matter in from infinity.}
\item{We find that energy conservation due to horizon expansion produces a situation loosely analogous to the classical electromagnetic interaction involved in chemical bond formation:  when an attractive interaction takes place as part of the bond formation, the increased magnitude of the negative potential energy is offset by an increase in kinetic energy so as to conserve the total energy. We identify this increase of kinetic energy with an increase in temperature and call the chemical interaction exothermic.  The case is not so simple with gravitational interaction, where gravity is associated with the metric. Rather than kinetic energy increase directly, in the sense of increased temperature or altered particle motions, we will invoke an analogue of the work-energy theorem, and use the result above that $p=-\rho$ to associate an energy change in the sense of $p\Delta V$, where $\Delta V$ is the consequence of the expansion of the metric.}
\item{We will work in the coordinate frame rather than in the standard observer (or conformal) frame. This simplifies things for us and makes possible a back of the envelope estimate of the spatial parts of the flat FRW metric.}
\item{We will discuss the earliest universe separately. The mechanism of the present model raises some new issues, suggesting a mechanism for matching the boundary conditions of the present era, well described with a flat FRW cosmology, with possibly alternative cosmologies of a prior evolution, provided the alternative cosmologies possess sufficiently high densities and short enough horizons in the relevant time frame.}
\end{itemize}

\section{\label{sec:kinetic}The kinetic response of the metric.}

We will treat the universe as if it were an ideal gas, and identify the change in energy density due to expansion with work in the sense of $W= -p\Delta V$. We know from $\ddot a \ne 0$ that there are pressures at work. Here the gravitational response is assumed to not be in terms of expansion at the speed of light, $c\Delta t$, but in terms of changes in the metric by way of $\dot a >0$. The volume rate of change at the horizon due to metric effects is therefore $4\pi t^2 \dot a$. This work is done against the pressure $p=-\rho$, which is constant, so we will identify ($c=1$)
\begin{equation}
\frac{\partial W}{\partial t} = -p \frac{\partial V}{\partial t} = \rho (4\pi t^2) \dot a
\label{dq:dotK}
\end{equation}
The fact that $\ddot a > 0$ confirms that a positive result is the correct choice, and this work might be analogized to stretching of a membrane. For comparison with the energy densities of matter, radiation, and whatever other energy densities may be under consideration, we need to convert this to a density:
\begin{equation}
\dot \rho_W = \left( {\frac{\partial W}{\partial t}} \right) \big/ \, \left( {\frac{4\pi }{3} t^3} \right)  = 3\rho \frac{\dot a}{t} 
\label{eq:rhoW}
\end{equation}

\section{\label{sec:potential}The effect of the expanding horizon on the gravitational potential.}

 The semi-classical gravitational mass/energy inside the horizon is 
\begin{equation}
m_{\mathrm{in}} = \frac {4\pi}{3} t^3 \rho
\end{equation}
and a shell  $4\pi t^2  \Delta t$ of density $\rho$ at radius $t$ should produce a change in potential energy
\begin{equation}
\Delta U = -G\left\{ \frac {4\pi}{3} t^3 \rho \right\} \,\,\left\{ ( 4\pi t^2 \rho ) \Delta t \right\}
\frac {1}{t}\, .
     \label{eq:dU}
\end{equation}
We thus conclude that
\begin{equation}
\frac{\partial U}{\partial t} =    - \frac{16 \pi^2 G}{3} t^4 \rho^2
     \label{eq:dotU}
\end{equation}
gives the time evolution of the gravitational potential due to the expansion of the horizon of that observer.

Similarly, we need to convert this to an energy density as well:
\begin{equation}
\dot\rho_U = \left( {\frac{\partial U}{\partial t}} \right) \big/\, \left( {\frac{4\pi }{3} t^3}\right)  =- 4  \pi G\rho^2 t
\label{eq:dotrhoU}
\end{equation}

\section{\label{sec:combined}The combined effect of horizon and response.}

Continuing in the spirit of $\dot \rho = 0$, we include consideration of the mass energy density, $\rho_m$ and the radiation energy density, $\rho_\gamma$,
\begin{equation}
\frac{\partial \rho}{\partial t} = \frac{\partial \rho_m}{\partial t} + \frac{\partial  \rho_\gamma}{\partial t} + \frac{\partial \rho_W}{\partial t} + \frac{\partial \rho_U}{ \partial t} +\cdots = 0 
     \label{eq:Econ1}
\end{equation}
precisely. We include the $\cdots$ because some readers may wish to include additional terms, such as a density due to neutrinos. Current wisdom is that the dominant term in the expression for $\rho$ has varied over time, first the radiation term, then the matter term, and presently the $\rho_W$ term, which we identify with Dark Energy, seems to dominate. A mechanism for the alteration of dominant terms is provided below.

Our present appraisal will adopt the standard conclusions that at all times the density of matter is proportional to $a^{-3}$ and the density of radiation is proportional to $a^{-4}$, $\rho_m = k_m a^{-3}$ and $\rho_\gamma = k_\gamma a^{-4}$. 

 The rule for conservation of energy becomes for the energy densities under immediate consideration becomes
\begin{eqnarray}
\dot \rho = &&   -\, 3 k_m \frac{\dot a}{a^4} - 4 k_\gamma \frac{\dot a}{a^5}  \nonumber \\
&& \,  - \,  4  \pi G \rho^2 t    \nonumber \\
&& + \,  3\rho \frac{\dot a}{t}  \nonumber \\
=&& 0 \,\, .
      \label{eq:econ2}
\end{eqnarray}
The quintic in$1/a$ suggests this is a very difficult differential equation to solve, and that it probably does not have an analytic solution. 

We can assume that the $\rho_W$ can be  identified with Dark Energy, whose energy density is estimated to have a current value of approximately $0.7 \rho_{cr} \approx 0.68 \times 10^{-29} g\, cm^{-3}$. The work of the expanding metric is the only positive contribution to the evolving total energy density. This energy density is not  associated with any cosmological constant or with any curvature. 

Note that the rate of change of the potential energy density depends on the constant total energy density of the universe, and time, so that the changes with time in relative distribution of the matter and radiation energy densities is ineveitable. The horizon must expand with the passage of time, and in consequence the spatial metric of the universe must expand,  so as to conserve energy through the mechanism of the work energy density. The inescapable expansion effects the densities of matter and radiation in the universe. This seems to be the primary cross-linking mechanism between the constituents of the universe and geometry, and govern their evolution over time.

\section{\label{sec:estimate}An estimate of the scale factor and the spatial metric}

Here we present a back of the envelope calculation which estimates current values of the scale factor $a$ and the spatial metric elements $a^2$. Taking $\rho = \rho_{cr} = \textrm{constant}$, using current values for $\rho_m \approx .3 \rho_{cr }a^{-3}$ and $\rho_\gamma \approx  2.47\times 10^{-5} h^{-2}a^{-4} \rho_{cr}$, $\rho_{cr} = 1.88 h^2 \times  10^{-29} \textrm{g cm}^{-3}$, $h\approx 0.72$, we can evaluate equation (\ref{eq:Econ1}).  Taking $t= 12.6 \times 10^9$ years ($\approx 3.97 \times 10^{17} \,\textrm{ s}$) and $H_o = 73 \, \textrm{km s}^{-1}\textrm{Mpc}^{-1}$, $\rho = \rho_{cr} \approx .97 \times 10^{-29} \, \textrm{g cm}^{-3}$,  $\dot a = H_o t$,
$\dot a \approx  2.82\times 10^{5} \, \textrm{km s}^{-1}= 2.82\times 10^{10} \textrm{cm s}^{-1}$ , and our quintic becomes 
\begin{eqnarray}
&&  0= -\, 3 (.3\rho)(c^2) \frac{\dot a}{a^4} \nonumber \\
&&\qquad  - 4 (2.47\times 10^{-5} \,h^{-2}\rho) (c^2)\frac{\dot a}{a^5}  \nonumber \\
&& \qquad  - \,  4  \pi G \rho^2 ct    \nonumber \\
&& \qquad + 3\rho (c^2)\frac{\dot a}{ct}  \nonumber \\
&&=\,\left\{ 5.37 \times 10^6 \frac{1}{a^5} 
\,  +\, 2.538\times 10^{10} \frac{1}{a^4} \right\} (c^2\rho)\nonumber \\
  &&\qquad +\,  9.72\times 10^{-8}\rho  -\,4.43\times 10^{-18} (c^2\rho)\,  \, .
\label{eq:quintic}
\end{eqnarray}
The $c^2$ and $G$ factors perform their respective mass-energy conversion so that we have common units, there is a common factor of $\rho$, and g-cm-s units were used throughout. The only physical root for $1/a$  (according to the Mathematica numeric solver) is $2.04\times 10^{-7}$, for a scale factor $a \approx 4.89 \times 10^6$ and a current value of the spatial metric element of $a^2 \approx 2.39\times 10^{13}$. Very very large, but the accelerating expansion has been going on for approximately $10^{10}$ years.

\section{\label{sec:primordial} In the earliest primordial universe}

There is today no credible predictive model of the very earliest universe. The common understanding of the earliest universe seems to the author to be summarized as follows: at the earliest times, there was some combined field, and after some interval the energy density of this combined field declined to a point that particles began to condense out of whatever had been before. Droplets condensing out of water vapor is a common analogy. Some scale threshold was crossed resulting in a condensation, and what emerged was a universe governed by physical rules much like we see today. Through condensation, the unified field had evolved into something like the Standard Model of particle physics  plus gravity, with gravity being described by General Relativity. Prior to the condensation, the physics of the unified fields was different from what we know today.

During and after the condensation era, we fully expect metric expansion to occur in response to the  expansion of the horizon as described preceding. We may, however,  anticipate additional effects if particle number densities are high and horizons are small at the time of first gravitational interaction between particles. Exactly what to expect will depend on a precise model of the condensation period, which we don't have, but it is easy to anticipate that the simplistic treatment above may not necessarily be terribly relevant during times close to the condensation era: even though the same basic mechanism of metric expansion driven by horizon expansion may be in play, fuller application of Einstein's equations may be necessary to account for all possible general relativistic effects. In addition, there are additional particle physics mechanisms which may come into play as a byproduct of large magnitude energy densities, should there have been sufficiently high number densities coupled with short distance interactions involving $1/r$ potentials. A ``hot-dense'' Big Bang may well have different physical processes  from a Big Bang which is merely ``hot''.

The Alcubierre metric with its warp bubbles~\cite{alcubierre} requires large magnitude negative energy densities, and while it is possible that stupendous energy densities (and possibly exotic matter) might be required to create macroscopic warp bubbles~\cite{pfenning}, more recent estimates suggest all that is actually required are merely very large magnitude energy densities~\cite{krasnikov}. E.g., if the first interactions of matter particles with attractive $1/r$ potentials should occur when horizons are of the order of a few Planck lengths, this may well generate the necessary large magnitude negative energy densities to produce significant quantities of these so-called warp bubbles. This would make possible not only superluminal expansion  of the primordial universe during the condensation era, but would make possible the adjustment of its energy density as well (by sending volumes of spacetime together with their contents ``elsewhere''). Thus, such interactions could provide a physical mechanism for adjusting the energy density towards some favored value which eliminates curvature energy in accordance with a metric minimum energy principle. In effect, flat cosmologies may be dynamically attractive while gravitational energy density adjustment mechanisms are active.

 Probably the most fruitful way of regarding Alcubierre warp bubbles and similar high material density related primordial processes is to consider them mechanisms for boundary matching, connecting the boundary of a cosmology existing prior to the condensation era with the boundary of the cosmology of the post condensation era (which persists even today). In the warp bubbles generated by high density and short range primordial interactions, we have at least one candidate for a mechanism of evolving a curved cosmology into a flat cosmology, given proper initial conditions.

\bibliography{flatfrw}

\begin{thebibliography}{9}
\bibitem[{not()}]{note1}
\bibinfo{note}{The flat FRW cosmology is a textbook standard, see, e.g.,
  ~\cite{weinberg} or ~\cite{dodelson}}.
\end{thebibliography}

\end{document}